\documentclass[aps,prd,reprint,onecolumn,superscriptaddress,nofootinbib,floatfix]{revtex4}
\pdfoutput=1
\usepackage{epsfig, subfigure}
\usepackage{setspace}
\usepackage{booktabs, tabularx} 
 
\usepackage{amsmath,bm,bbm}
\usepackage{hyperref}
\usepackage{graphicx}
\usepackage{enumerate}
\usepackage{dcolumn}
\usepackage{bm}
\usepackage{longtable}
\hypersetup{
    pdfnewwindow=true,      
    colorlinks=true,       
    linkcolor=blue,          
    citecolor=blue,        
    filecolor=blue,      
    urlcolor=blue           
}

\def\lsim{\lesssim}  
\def\gsim{\gtrsim}

\newcommand{\beq}{\begin{equation}}  
\newcommand{\eeq}{\end{equation}}
\newcommand{\bea}{\begin{eqnarray}}  
\newcommand{\eea}{\end{eqnarray}}
\newcommand{\epm}{\ensuremath{e^{\pm}\;}}

\newcommand{\sigv}{\ensuremath{\langle \sigma v\rangle}}
\newcommand{\sigvchi}{\ensuremath{\langle \sigma v\rangle_{\chi}}}
\newcommand{\sigvdm}{\ensuremath{\langle \sigma v\rangle_{\rm DM}}}
\newcommand{\sigvcmb}{\ensuremath{\langle \sigma v\rangle_{\rm CMB}}}
\newcommand{\sigvann}{\ensuremath{\langle \sigma v\rangle_{\rm ann}}}
\newcommand{\omchi}{\ensuremath{\Omega_{\chi}}}
\newcommand{\omdm}{\ensuremath{\Omega_{\rm DM}}}

\newcommand{\omhchi}{\ensuremath{\Omega_{\chi}{\rm h}^{2}}}
\newcommand{\omhdm}{\ensuremath{\Omega_{\rm DM}{\rm h}^{2}}}

\newcommand{\mchi}{\ensuremath{m_\chi}}
\def\3he{$^3$He}
\def\4he{$^4$He}
\def\7li{$^7$Li}

\newcommand{\ie}{{\it i.e.}}
\newcommand{\eg}{{\it e.g.}}

\begin{document}

\title{\mbox{\hspace{-0.5cm}CMB Constraints On The Thermal WIMP Mass And Annihilation Cross Section}}

\author{Gary Steigman}
\email{steigman.1@osu.edu}
\affiliation{\mbox{Center for Cosmology and AstroParticle Physics, Ohio State University,}}
\affiliation{Department of Physics, Ohio State University, 191 W.~Woodruff Ave., Columbus,  43210 OH, USA}

\date{\today}

\begin{abstract}
A thermal relic, often referred to as a weakly interacting massive particle (WIMP), is a particle produced during the early evolution of the Universe whose present (relic) abundance depends only on its mass and its thermally averaged annihilation cross section (annihilation rate factor) \sigvann.  Late time WIMP annihilation has the potential to affect the cosmic microwave background (CMB) power spectrum. Current observational constraints on the absence of such effects provide bounds on the mass and the annihilation cross section of relic particles that may, but need not be dark matter candidates.  For a WIMP that is a dark matter candidate, the CMB constraint sets an upper bound to the annihilation cross section, leading to a lower bound to its mass that depends on whether or not the WIMP is its own antiparticle.  For a self-conjugate WIMP, $m_{\rm min} = 50\,f\,{\rm GeV}$, where $f \leq 1$ is an electromagnetic energy efficiency factor.  For a non self-conjugate WIMP, the minimum mass is a factor of two larger.  For a WIMP that is a subdominant component of the dark matter density there is no bound on its mass and the upper bound to its annihilation cross section imposed by the CMB transforms into a lower bound to its annihilation cross section.  These results are outlined and quantified here using the latest CMB constraints for a stable, symmetric (equal number of particles and antiparticles), WIMP whose annihilation is s-wave dominated, and for particles that are, or are not, their own antiparticle.
\end{abstract}

\maketitle

\section{Introduction}
\label{sec:intro}

Any new, massive, beyond the standard model particle, $\chi$, populated by interactions with the standard model particles present during the early evolution of the Universe and in thermal equilibrium with them, will annihilate when it becomes non-relativistic (for temperatures $T < \mchi$), reducing its abundance (relative to, \eg, photons).  Initially, the annihilation rate (per particle) $\Gamma_{\rm ann} = n_{\chi}\sigvann \propto T^{3}$, exceeds the expansion rate, the Hubble parameter, $H \propto T^{2}$, and $n_{\chi} = n_{\chi,eq}$, the number density in thermal equilibrium at temperature $T$.  Annihilations continue until the annihilation rate (per particle) becomes too slow compared with the expansion rate of the Universe, and the abundance ``freezes out", when $T = T_{f}$, where $\Gamma_{{\rm ann},f} = H_{f}$.  For $T < T_{f}$, $\Gamma_{\rm ann} < H$, preserving the relic number of particles in a comoving volume.  For symmetric ($n_{\chi} = n_{\bar{\chi}}$), stable particles, the late time abundance (\eg, the ratio of the current mass density to the present critical mass density, $\omchi \equiv (\rho_{\chi}/\rho_{\rm crit})_{0}$) is determined by the annihilation cross section, $\sigvchi = \sigvann$\,\footnote{In the discussion here, \sigvchi~always refers to the s-wave annihilation rate factor, \sigvann, and \sigvchi~and \sigvann~are used interchangeably.}.  Such particles are referred to here as thermal relics or, interchangeably, as weakly interacting massive particles (WIMPs).  In the WIMP ``paradigm", the annihilation cross section is chosen ($\sigvchi = \sigvdm$) so that the WIMP accounts for the observationally inferred dark matter (DM) density, $\omchi = \omdm$.

However, annihilations don't cease when $T = T_{f}$.  It is important to realize that even though when $T < T_{f}$, $\Gamma_{\rm ann} < H$, the WIMP continues to annihilate throughout the evolution of the Universe, independent of whether or not it is a dark matter candidate.  Although long after freeze out (when $T \ll T_{f} \ll \mchi$) residual annihilations are rare ($\Gamma_{\rm ann} \ll H$), the energy injected from each annihilation, $2\,\mchi$, may be large compared to the thermal energy, leading to observational constraints on the late time annihilations.  Some of the earliest constraints on late time annihilation have come from big bang nucleosynthesis (BBN) \cite{reno,hagelin,fkt,osw}.  The BBN constraints were supplemented (superseded) in a series of papers \cite{sz,Burigana,husilk,msw} using constraints on the deviation from black body of the spectrum of the cosmic microwave background radiation (CMB) frequency spectrum \cite{firas} and from the reionization and heating of the intergalactic gas \cite{cirelli}.  For a discussion and references to complementary constraints from laboratory experiments, the effect on stellar evolution, from gamma ray observations of the Galaxy and beyond, see, \eg, \cite{msw}.  At present, the Planck satellite observations of the CMB power spectrum \cite{planck} appears to offer the most restrictive constraints on late time annihilations and, in particular, on the annihilation cross section, \sigvann~\cite{padman,galli09,galli11,hutsi,natarajan,slatyer}.  For a dark matter candidate (\ie, if \omchi~= \omdm), the Planck CMB observations, by requiring that $\sigvann = \sigvchi = \sigvdm \leq \sigvcmb \propto \mchi$, set a {\bf lower} bound to the $\chi$ mass, $m_{\rm min}$, in the interesting range of a few tens of GeV (see, \eg, \cite{slatyer}).  The most recent Planck results \cite{planck15} improve on these results, increasing the value of $m_{\rm min}$ by more than a factor of two.

In this paper these constraints from the CMB on \mchi~and $\sigvchi = \sigvann$ are revisited, noting that there are different constraints depending on whether the thermal relic particle is, or is not, its own antiparticle ($\chi = \bar{\chi}$ or $\chi \neq \bar{\chi}$), and investigating how the constraints change if the WIMP is only a subdominant component of the dark matter ($\omchi < \omdm$).  In particular, it is noted here that if $\omchi < \omdm$, then the lower bound on the WIMP mass from the CMB disappears, and the upper bound on $\sigvann = \sigvchi$ becomes a lower bound, $\sigvchi \geq \sigv_{\rm min} \equiv (\sigvdm)^{2}/\sigvcmb$, where \sigvdm~is the value of the annihilation cross section, \sigvchi, required for $\omchi = \omdm$.  An overview of the calculation of the thermal relic abundance and its connection with the particle mass and annihilation cross section is presented in \S\,\ref{sec:sigv}, as a prelude to considering in \S\,\ref{sec:cmb} the relation between the annihilation cross section $\sigv_{\chi}$ and the CMB cross section $\sigv_{\rm CMB}$ that emerges from the CMB constraint on late time annihilations at the epoch of recombination.  The results presented here are summarized and discussed in \S\,\ref{sec:concl}.

\section{The Thermal Relic Annihilation Cross Section And The Relic Abundance}
\label{sec:sigv}

In this paper, a WIMP is a particle whose relic abundance is determined by its thermally averaged annihilation cross section (annihilation rate factor, \sigvann).  This excludes from consideration here any particle -- antiparticle asymmetry ($n_{\chi} \neq n_{\bar{\chi}}$), since in this case the relic abundance is determined by the adopted asymmetry (\eg, by the chemical potential) and not directly by the annihilation rate factor.  Also excluded from consideration here is a long-lived, unstable particle whose late time energy injection depends on its mass and lifetime.  This latter case has been analyzed in \cite{kamionkowski}.  

For a thermal relic, during the early evolution of the Universe when the temperature exceeds the WIMP mass, annihilations are compensated by the inverse processes (particle creation), and the WIMP abundance is preserved as the Universe expands and cools.  In this early, extremely relativistic (ER) regime, the ratio of $\chi$ particles to photons, $n_{\chi}/n_{\gamma} = n_{\chi,eq}/n_{\gamma} = {\rm constant}$ since, in this ER regime, $n_{\chi,eq} \propto T^{3} \propto n_{\gamma}$.  When $T \lsim m_{\chi}$, particle creation is suppressed by an energy barrier and the thermal relic abundance decreases exponentially ($n_{\chi,eq}/n_{\gamma} \propto x^{3/2}e^{-x}$, where $x \equiv m_{\chi}/T$).  Nonetheless, depending on the strength of the coupling between the $\chi$ and the standard model particles whose masses are $< T$, in this intermediate non-relativistic (NR) regime, $\Gamma_{\rm ann} \gg H$, and the thermal relic abundance remains close to equilibrium, $n_{\chi} \approx n_{\chi,eq}$, decreasing exponentially as the Universe expands and cools.  This continues until the inverse reactions become too slow to maintain equilibrium (\ie, when the deviation from equilibrium, $(n_{\chi} - n_{\chi,eq})/n_{\chi,eq}$, grows to of order unity, at $T = T_{*}$, when $(n_{\chi,eq}/n_{\gamma})_{*} \propto x_{*}^{3/2}e^{-x_{*}} \ll 1$).  Thereafter, for $T < T_{*}$, annihilations dominate, quickly reducing, even further, the abundance of the $\chi$ particles, until the annihilations finally ``freeze out" at $T = T_{f} \approx T_{*}/2$, when $\Gamma_{{\rm ann},f} = (n_{\chi}\sigvann)_{f} = H_{f}$.  For a thermal relic, the WIMP abundance (\eg, the ratio of the current WIMP mass density to the present critical mass density, $\omchi \equiv (\rho_{\chi}/\rho_{\rm crit})_{0}$) is determined by the annihilation cross section, $\sigvchi = \sigvann$.

The time (temperature) evolution of a symmetric ($n_{\chi} = n_{\bar{\chi}}$), stable, thermal relic follows the evolution equation first described by Zeldovich \cite{zeldovich},
\beq
dN_{\chi}/dt = dN_{\bar{\chi}}/dt = \sigvann\,(n_{\chi,eq}^{2} - n_{\chi}\,n_{\bar{\chi}})V = \sigvann\,(n_{\chi,eq}^{2} - n_{\chi}^{2})V\,,
\eeq
where $N_{\chi} = n_{\chi}V$ is the number of $\chi$ particles in the comoving volume $V$ at time $t$ and $n_{\chi}$ is their number density, and similarly for the antiparticle $\bar{\chi}$.  For the analysis presented here, attention is focused on a relic particle whose annihilation is s-wave dominated\,\footnote{At late times and low temperatures (\eg, around recombination), p-wave annihilation is suppressed by a factor of $T/\mchi$ (or $(v/v_{f})^{2}$) that is typically $\ll 1$, leading to much weaker CMB constraints for this case.  ``Sommerfeld" or resonance enhancement (or suppression) of the annihilation cross section is model dependent and is ignored in the discussion here.}.  For this choice, the annihilation rate factor (``cross section"), $\sigvann \equiv \sigvchi$, is independent of temperature or velocity, resulting in a cross section that is the same at freeze out and during all subsequent epochs in the evolution of the Universe as well as in all environments in the present Universe (\eg, the Sun, galaxies, clusters of galaxies, etc.).  The thermal relic evolution equation, a form of the Riccati equation, cannot be integrated in closed form, but needs to be solved either numerically or by using an analytic approximation first outlined by Zeldovich \cite{zeldovich}.  The quantity that is actually predicted by the numerical and semi-analytic solutions of the evolution equation is the product of three terms: the mass, the annihilation cross section, and the frozen out abundance of the relic particles.  The latter may be written in terms of the present ($z = 0$) number density $n_{\chi0} \equiv n_{\chi} = n_{\bar{\chi}}$, so that post freeze out, for $z \leq z_{f}$, where $z_{f}$ corresponds to $T = T_{f}$, $n_{\chi}(z) = n_{\chi0}(1 + z)^{3}$.  The solution to the evolution equation predicts the combination $A(m_{\chi}) \equiv m_{\chi}n_{\chi0}\sigvchi$ (see, \eg,\,\cite{zeldovich} or \cite{gs79}), where $A(\mchi)$ is a slowly varying function of the mass (see, \eg,\,\cite{gg,sdb}).

\begin{figure}[!t]
\begin{center}
\includegraphics[width=0.5\columnwidth]{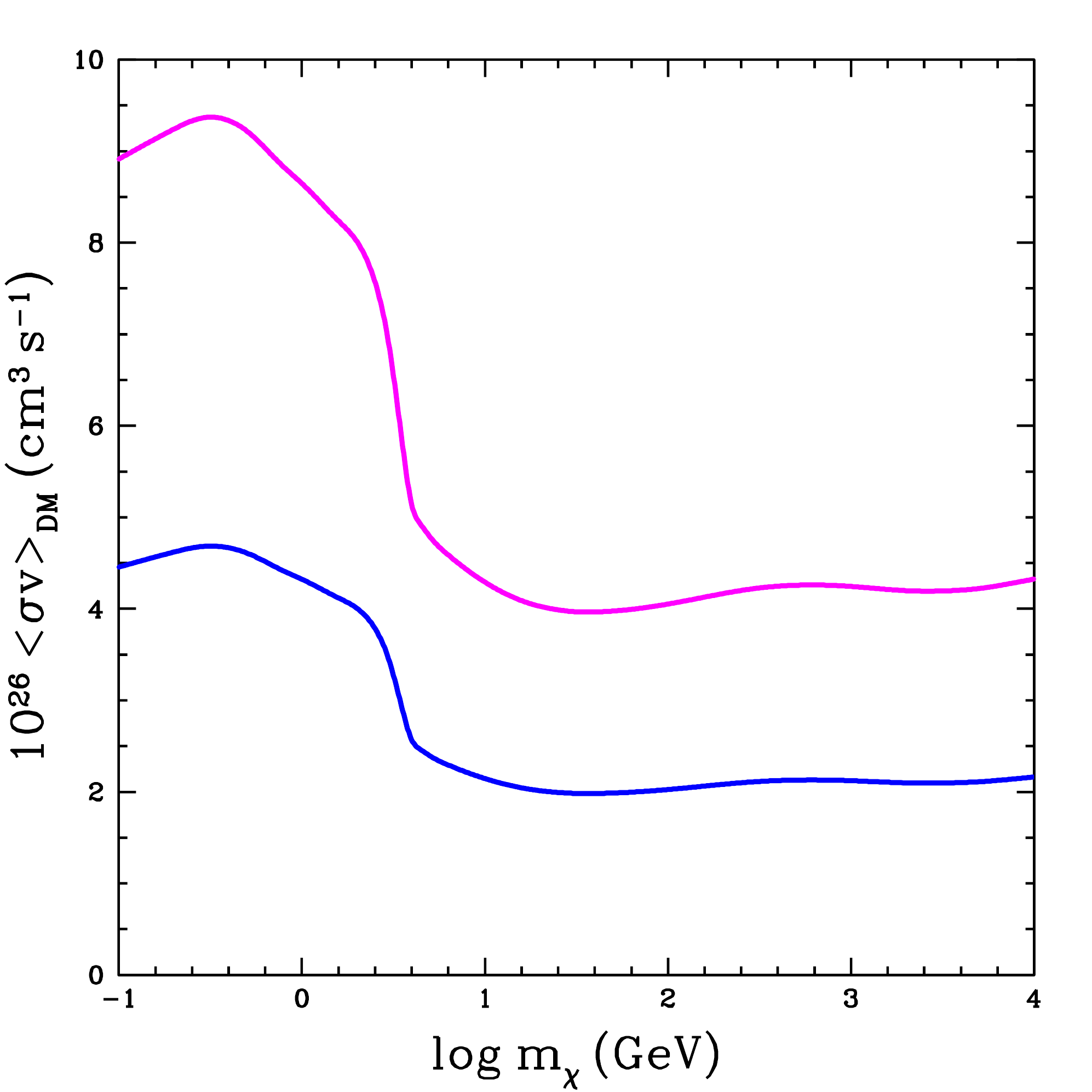}
\caption{(Color online) The s-wave, thermal annihilation cross section (rate factor), $\sigvdm = \sigvann \equiv \sigvchi$, as a function of the relic particle mass, \mchi, required for the WIMP to account for the observed dark matter density ($\omhchi = \omhdm = 0.12$).  The top (purple) curve is  for a non self-conjugate ``NS" particle ($\chi \neq \bar{\chi}$) and the bottom (blue) curve is for a self-conjugate ``S" particle ($\chi = \bar{\chi}$).    Only along the curves does the WIMP account for the DM ($\omchi = \omdm$).  Although the regions above the curves are allowed, in these regions the WIMP is a subdominant contributor to the DM ($\omchi < \omdm$).  The regions below the curves are forbidden since, in these regions the relic WIMP mass density exceeds the DM density ($\omchi > \omdm$).}
\label{fig:sigvdm14}
\end{center}
\end{figure}

The quantitative relation between the annihilation cross section and the relic mass density depends on whether or not the particle is its own antiparticle ($\chi = \bar{\chi}$ or $\chi \neq \bar{\chi}$).  When $\chi = \bar{\chi}$, the particle will be referred to as ``self-conjugate" (S), while for the case where $\chi \neq \bar{\chi}$ the particle will be referred to as ``non self-conjugate" (NS).  To fix the normalization for the $\sigvchi - \omhchi$ relation\,\footnote{It is convenient (and conventional) to parameterize the present mass density by \omhchi, where $\omchi = \rho_{\chi}/\rho_{\rm crit}$, $\rho_{\rm crit}$ is the present Universe critical mass density, and the present value of the Hubble parameter is H$_{0} = 100\,{\rm h}\,{\rm km\,s^{-1}\,Mpc^{-1}}$.}, a choice needs to be made.  For S particles the mass and number densities are related by $\rho_{\chi{\rm S}} = m_{\chi}n_{\chi{\rm S}}$, while for NS particles $\rho_{\chi{\rm NS}} = m_{\chi}(n_{\chi{\rm NS}} + n_{\bar{\chi}{\rm NS}}) = 2\,m_{\chi}n_{\chi{\rm NS}}$.  As a result, there are two possible choices.  For example, it might be useful to choose to have the relic {\bf number} densities equal, independent of whether or not the particle is its own antiparticle.  In this case, for $n_{\chi{\rm S}} = n_{\chi{\rm NS}}$, the solution of the evolution equation (evaluated at the same mass) requires that $\sigv_{\rm N} = \sigv_{\rm NS}$.  For this choice the relic {\bf mass} densities differ, $\rho_{\chi\rm NS} = 2\,\rho_{\chi\rm S}$.  However, the more common choice is that, independent of whether the relic particle is or is not its own antiparticle, the relic {\bf mass} densities are the same, $\rho_{\chi{\rm NS}} \equiv \rho_{\chi{\rm S}}$.  In this case, $\sigv_{\rm NS} = 2\,\sigv_{\rm S}$, and the abundances (\ie, the post freeze out number densities) of the thermal relics differ, $n_{\chi{\rm S}} = 2\,n_{\chi{\rm NS}}$.

For the choice made here, $\Omega_{\rm NS} = \Omega_{\rm S}$, the results for $\sigvchi = \sigvdm$ are shown by the two curves (for S and NS relic particles) in Figure \ref{fig:sigvdm14} for $\omhchi = \omhdm = 0.12$ \cite{planck}.  As may be seen in Figure \ref{fig:sigvdm14}, for a fixed value of the present mass density contributed by a thermal relic ($\chi$), the required cross section is {\bf not}, independent of the particle mass, as is often assumed/claimed.  The results shown here (and earlier in reference \cite{sdb}) are in quantitative agreement with those that follow from the publicly available DarkSUSY code \cite{darksusy} (P. Gondolo, Private Communication).

After freeze out ($z \leq z_{f}$), the number density of relic particles varies with the redshift as $(1 + z)^{3}$, as does the number density of baryons (the relic of an asymmetry between baryons and antibaryons in the early Universe), so that for $z < z_{f}$, $n_{\chi}/n_{\rm B} = {\rm constant}$.  Post freeze out, the annihilation rate per relic particle is $\Gamma_{\rm ann} \equiv n_{\chi}\sigvchi = A(\mchi)/\mchi$, so that when evaluated at the same mass, $\Gamma_{\rm ann,NS} = \Gamma_{\rm ann,S}$.  However, the late time annihilation rates per unit volume differ between S and NS thermal relics.  The annihilation rate per unit volume is $n_{\chi}\Gamma_{\rm ann} = n_{\chi}A(\mchi)/\mchi$, so that,
\beq
{(n_{\chi}\Gamma_{\rm ann})_{\rm NS} \over (n_{\chi}\Gamma_{\rm ann})_{\rm S}}  = {n_{\chi{\rm NS}} \over n_{\chi{\rm S}}} = {1 \over 2}\bigg({\rho_{\chi{\rm NS}} \over \rho_{\chi{\rm S}}}\bigg)\,.
\eeq
For the choice adopted here, $\rho_{\chi{\rm NS}} \equiv \rho_{\chi{\rm S}}$, requiring $\sigv_{\rm NS} = 2\sigv_{\rm S}$, this leads to $(n_{\chi}\Gamma_{\rm ann})_{\rm S} = 2 (n_{\chi}\Gamma_{\rm ann})_{\rm NS}$.  However, since in each annihilation, an energy $E = 2\mchi$ is injected (the same for S and NS), the energy injection rate per unit volume, $d\epsilon/dt = 2\mchi(n_{\chi}\Gamma_{\rm ann})$.  But, provided that $\rho_{\chi{\rm NS}} \equiv \rho_{\chi{\rm S}}$, $(d\epsilon/dt)_{\rm NS} = \rho_{\chi{\rm NS}}\Gamma_{\rm ann,NS} = \rho_{\chi{\rm S}}\Gamma_{\rm ann,S} = (d\epsilon/dt)_{\rm S}$.  Thus, although the late time annihilation rates per unit volume (or, per baryon) differ between S and NS particles, the energy injection rates per baryon are the same for both cases.

The results for the dependence of the annihilation cross sections on the WIMP mass shown in Figure\,\ref{fig:sigvdm14}, spanning the mass range $100\,{\rm MeV} \leq \mchi \leq 10\,{\rm TeV}$\,\footnote{Here, masses and mass densities are measured in energy units ($c \equiv 1$).}, are for the case where the WIMP accounts for the  dark matter ($\omchi = \omdm$, where $\omhdm = 0.12$ \cite{planck}).  Only those WIMPs whose s-wave annihilation cross section and mass lie on one or the other of the two curves in Fig.\,\ref{fig:sigvdm14} can account for the observationally inferred dark matter mass density.  These curves should have a finite thickness, not shown in the figure, to account for the theoretical uncertainty in the \sigvdm~-- WIMP mass relation ($\sim 5 - 10 \%$ \cite{sdb}) and the uncertainty (smaller) in \omhdm~($\sim 1.5 - 2 \%$ \cite{planck15}).  As may be seen from Fig.\,\ref{fig:sigvdm14}, for either kind of relic particle, S or NS, \sigvdm~is very nearly constant (independent of mass) for $\mchi \gsim 10\,{\rm GeV}$, with the ratio of cross section values at the same mass scaling with the factor of two explained above.  However, for $\mchi \lsim 10\,{\rm GeV}$, the required annihilation cross section, \sigvdm, increases by a factor of two or more, a reflection of the changing number of degrees of freedom contributing to the energy and entropy densities as the relic particle is freezing out (see, \eg, the discussion in \cite{sdb}).  As noted above, the quantity predicted by the numerical and semi-analytic solutions of the evolution equation that lead to the results shown in Fig.\,\ref{fig:sigvdm14}, is the product of the annihilation rate factor \sigvchi~and the mass density parameter, \omhchi, (see, \eg,\,\cite{zeldovich} or \cite{gs79}).  In general, for $\omchi \neq \omdm$ (but for \omhchi~not too different from \omhdm), $\sigvchi(\omhchi) \approx \sigvdm(\omhdm)$.  When \omhchi~differs significantly from \omhdm, there is a logarithmic correction to this approximation \cite{sdb},
\beq
\bigg({\sigvchi \over \sigvdm}\bigg)\,\bigg({\omhchi \over \omhdm}\bigg) \approx 1 - {1 \over 8}\,{\rm log}\,\bigg({\omhchi \over \omhdm}\bigg) \approx 1 + {1 \over 8}\,{\rm log}\,\bigg({\sigvchi \over \sigvdm}\bigg)\,.
\label{eq:chivsdm}
\eeq

For values of \sigvchi~and \mchi~along the curves in Fig.\,\ref{fig:sigvdm14} ($\sigvann = \sigvchi = \sigvdm$), $\omchi = \omdm$, while the regions above (below) the curves correspond to, $\omchi < \omdm$ ($\omchi > \omdm$).  The regions below the curves (for each kind of particle) are excluded, since in these regions $\omchi > \omdm$.  The regions above the curves are allowed, but the particles corresponding to these combinations of cross section and mass will not account for the dark matter, but will only be subdominant contributors to the total dark matter mass density ($\omchi < \omdm$).

\section{The CMB Constraint On Energy Injection From Late Time Annihilations}
\label{sec:cmb}

After the WIMP has frozen out at redshift $z_{f}$,  its number density redshifts, $n_{\chi}(z) = (1 + z)^{3}n_{\chi0}$, so that at redshift $z \leq z_{f}$ the annihilation rate per unit volume is $n_{\chi}(z)^{2}\sigvchi = (1 + z)^{6}n_{\chi0}^{2}\sigvchi$.  Since the relic number density may be written in terms of the relic mass density and the mass, and the relic mass density may be written in terms of the mass density parameter \omhchi, the annihilation rate per unit volume at redshift $z < z_{f}$ is proportional to the combination $(1 + z)^{6}(\omhchi/\mchi)^{2}\sigvchi$.  Assuming that none of the annihilation energy goes into neutrinos or the dark sector, the effect on the background plasma of late time annihilations at redshift $z$ due to the electromagnetic energy injection rate per unit volume is proportional to
\beq
d\epsilon_{\chi}/dt \propto(1 + z)^{6} (\omhchi)^{2}f(z)\sigvchi/\mchi\,,
\eeq
where $f(z)$ is an electromagnetic energy efficiency factor.  This expression is general in the sense that there is no assumption here that the WIMP is the dark matter.

\begin{figure}[!t]
\begin{center}
\includegraphics[width=0.5\columnwidth]{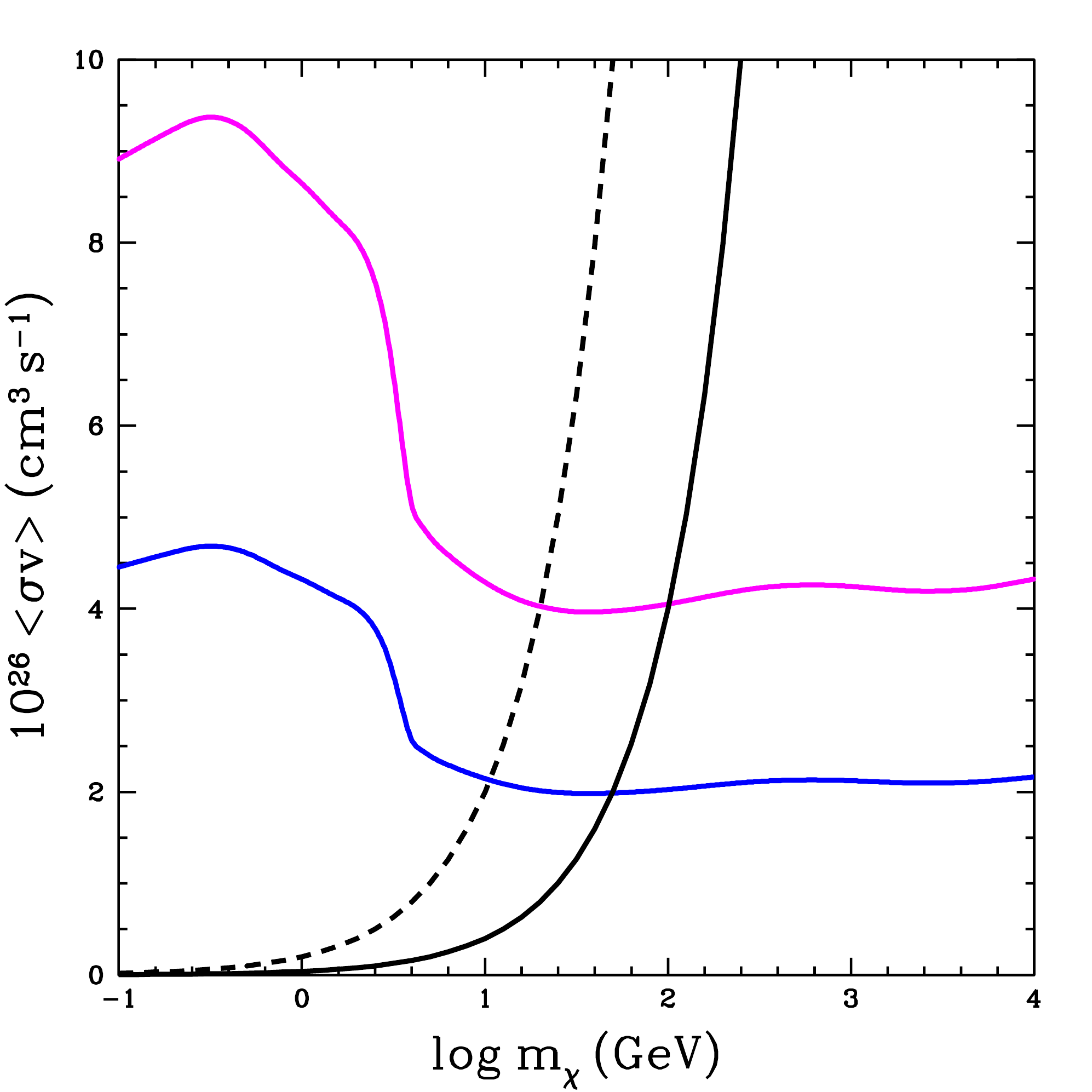}
\caption{(Color online) The annihilation rate factors $\sigvchi = \sigvdm$ from Fig.\,\ref{fig:sigvdm14} for a WIMP that is the DM are shown as functions of the WIMP mass, along with the CMB constraints on late time annihilations, \sigvcmb~(shown in black) for $f = 1$ (solid) and $f = 0.2$ (dashed).  The energy injection from late time annihilations is constrained by \sigvcmb~in that for $\omchi = \omdm$, the total annihilation cross section satisfies $\sigvchi = \sigvdm \leq \sigvcmb$, leading to a lower bound to the WIMP mass where the curves cross.  But, for $\omchi < \omdm$, there is no constraint on the WIMP mass and $\sigvchi < \sigvcmb(\omchi/\omdm)^{2}$.  See \S\,\ref{sec:ndm} for details.}
\label{fig:sigvallchi14bb}
\end{center}
\end{figure}

For particles in the mass range considered here, the relic abundance is frozen out at temperatures above a few MeV, well before recombination, while attention here is focussed on the energy injected at the recombination epoch ($z_{\rm rec} \approx 1100$), where $T_{\rm rec} \approx 0.3\,{\rm eV} \ll \mchi$.  Because the electromagnetic energy injected during this epoch has the potential to affect the CMB power spectrum\,\cite{padman,galli09,galli11,hutsi,slatyer,planck15}, the observed CMB power spectrum sets a constraint on late time annihilations, leading to a constraint on the combination $(\omhchi)^{2}f(z)({\sigvchi/\mchi})$ \cite{padman,galli09,galli11,hutsi,slatyer,planck15}.

Since the holy grail of particle cosmology is to identify a particle that is a potential dark matter candidate, in the literature it is almost always assumed that the WIMP is the DM ($\omchi = \omdm$), so that $(\omhchi)^{2}f(z)({\sigvchi/\mchi}) = f(z)\sigvdm/\mchi$.  This latter combination is usually identified as $p_{\rm ann} \equiv f(z)\sigvdm/\mchi$, and the CMB observations are used to set an upper bound to $p_{\rm ann}$.  It is convenient to introduce a CMB cross section (rate factor) related to $p_{\rm ann}$ by, $\sigvcmb \equiv \mchi\,p_{\rm ann}/f(z)$.  For the current Planck \cite{planck15} results, $\sigvcmb = 4\times 10^{-28}\,\mchi/f\,$ (in units of ${\rm cm^{3}\,s^{-1}}$, with \mchi~in GeV).  This cross section is shown as a function of \mchi~by the black curves in Figure \ref{fig:sigvallchi14bb} for two choices of $f$.  Note that while in general the efficiency factor is redshift dependent, the quantity that enters here is an ``effective" efficiency factor, $f \approx f(z = 600)$ (see \cite{slatyer, galli11} for discussion and further references).  According to \cite{slatyer}, it is likely that $0.2 \lsim f \lsim 1$ and these two choices are shown by the solid ($f = 1$) and dashed ($f = 0.2$) curves in Fig.\,\ref{fig:sigvallchi14bb}.

A naive, albeit incorrect interpretation of Fig.\,\ref{fig:sigvallchi14bb} would be that only the wedge-shaped regions {\bf above} the $\sigv_{\rm DM}$ curves and {\bf below} the $\sigv_{\rm CMB}$ curves are allowed for consistency with the mass density constraint, $\Omega_{\chi} \leq \Omega_{\rm DM}$, and with the CMB constraint on late time annihilations.  Indeed, for the case where the WIMP is the DM ($\omchi = \omdm$), there is a {\bf lower} limit to the WIMP mass, identified in Fig.\,\ref{fig:sigvallchi14bb} by the values of the masses at the crossings of the \sigvchi~and the \sigvcmb~curves.  For a self-conjugate WIMP, the minimum mass is $m_{\rm min} \approx 50\,(10)\,{\rm GeV}$ for $f = 1\,(0.2)$, and the minimum masses are twice as large for a non self-conjugate WIMP.  A stable, symmetric WIMP that accounts for the DM, whose annihilation is s-wave dominated, must have $\mchi \geq m_{\rm min}$ if it is to be consistent with the CMB.  If future CMB observations should reduce the current upper bound on \sigvcmb~\cite{planck15}, the lower bound on the minimum mass of a dark matter candidate will increase.  However, as explained below in \S\,\ref{sec:ndm}, if the WIMP does not account for all of the DM and $\Omega_{\chi} < \Omega_{\rm DM}$ is allowed, there is {\bf no} restriction on the WIMP mass and the only restriction on the annihilation cross section is that it {\bf exceed} some minimum value ($\sigv_{\chi} \geq \sigv_{\rm min}$).

\subsection{Constraints On A WIMP That Is Not The Dark Matter ($\omchi < \omdm$)}
\label{sec:ndm}

As  seen in Fig.\,\ref{fig:sigvallchi14bb}, if the WIMP is the DM, the CMB constraint on $p_{\rm ann}$, requiring that $\sigvchi = \sigvdm \leq \sigvcmb$, leads to a {\bf lower} bound to the WIMP mass.  However, allowing for a WIMP that is only a subdominant contributor to the DM mass density ($\omchi \leq \omdm$), the CMB constraint on $p_{\rm ann}$ leads to a constraint on the WIMP annihilation cross section that not only depends on \sigvcmb, but also on the ratio \omdm/\omchi,
\beq
\sigvchi \leq \sigvcmb\bigg({\omhdm \over \omhchi}\bigg)^{2}\,.
\label{eq:sigvcmb}
\eeq

The CMB constraint does not limit the mass of a WIMP whose contribution to the DM is subdominant, but it does set a {\bf lower} bound to the annihilation cross section for such a particle. Up to the logarithmic corrections in Eq.\,\ref{eq:chivsdm} (included later), the {\bf upper} bound to the annihilation cross section from the CMB, translates into a {\bf lower} bound to the annihilation cross section, as may be seen by combining Eqs.\,\ref{eq:chivsdm}\,\&\,\ref{eq:sigvcmb},
\beq
\label{eq:sigvmin}
\sigvchi \geq \sigv_{\rm min} \equiv {(\sigv_{\rm DM})^{2} \over \sigv_{\rm CMB}}\,\ \ \ {\rm or}\,\ \ \ {\sigvchi \over \sigvdm} \geq {\sigv_{\rm min} \over \sigvdm} \equiv {\sigv_{\rm DM} \over \sigv_{\rm CMB}}\,.
\eeq
For $\mchi \geq m_{\rm min}$, all values of $\sigvchi \geq \sigvdm$ are allowed since they correspond to $\omchi \leq \omdm$, and they also satisfy the CMB constraint (Eq.\,\ref{eq:sigvcmb})\,\footnote{For $\mchi \geq m_{\rm min}$, the CMB constraint may be rewritten as $\sigvchi(\omhchi/\omhdm)^{2} \approx  \sigvdm(\omhchi/\omhdm) \lsim \sigvdm \leq \sigvcmb$ (see Fig.\,\ref{fig:sigvallchi14bb} in support of this last inequality for $\mchi \geq m_{\rm min}$).}.  Another consequence of allowing $\omchi < \omdm$ is that the lower bound on the WIMP mass disappears.  For all $\mchi < m_{\rm min}$, $\sigv_{\rm min} = (\sigvdm)^{2}/\sigvcmb$, and all values of $\sigvchi \geq \sigv_{\rm min}$ are allowed since they, too, are consistent with the CMB constraint (Eq.\,\ref{eq:sigvmin}).

\begin{figure}[!t]
\begin{center}
\includegraphics[width=0.5\columnwidth]{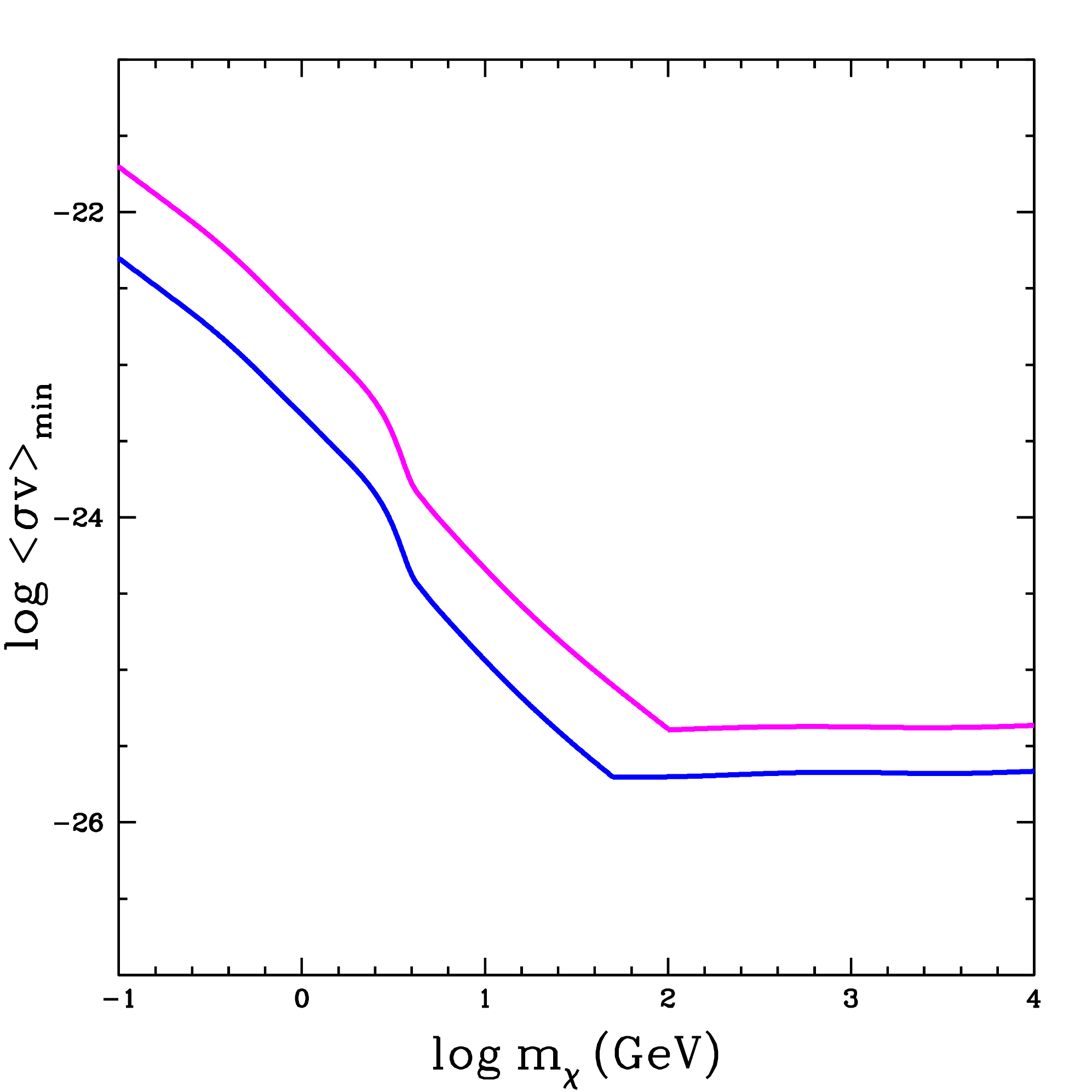}
\caption{(Color online) The minimum annihilation rate factor consistent with the CMB constraint for $f = 1$ and the mass density constraint $\omchi \leq \omdm$ is shown as a function of the WIMP mass.  The regions above the curves, corresponding to $\omchi \leq \omdm$, are allowed, while the regions below the curves, corresponding to $\omchi > \omdm$, are excluded.  The upper (purple) curve is for an NS WIMP ($\chi \neq \bar{\chi}$)  and the lower (blue) curve is for an S WIMP ($\chi = \bar{\chi}$).  For $\mchi \geq m_{\rm min}$, $\sigv_{\rm min} = \sigvdm$, while for $\mchi < m_{\rm min}$, $\sigv_{\rm min} = (\sigvdm)^{2}/\sigvcmb > \sigvdm$, corresponding to $\omchi < \omdm$.  See the text for details.}
\label{fig:sigvlogall14b}
\end{center}
\end{figure}

These results are shown in Figure \ref{fig:sigvlogall14b} for S and NS thermal relics, where $\sigv_{\rm min}$ (see Eq.\,\ref{eq:sigvmin}) is shown as a function of \mchi~(for $f = 1$).  For an NS particle, for all masses $\mchi \geq m_{\rm min} = 100\,{\rm GeV}$, $\sigv_{\rm min} = \sigvdm \approx 4\times 10^{-26}\,{\rm cm^{3}\,s^{-1}}$.  For masses below this value, $\sigv_{\rm min}$ increases with decreasing WIMP mass approximately as $\mchi^{-1}$, as seen in Figs.\,\ref{fig:sigvdm14}\,\&\,\ref{fig:sigvallchi14bb}, modulo the variation of \sigvdm~with mass shown in Fig.\,\ref{fig:sigvdm14} and the logarithmic corrections from Eq.\,\ref{eq:chivsdm} described below.  For an S particle, the same behavior for $\sigv_{\rm min}$ as a function of \mchi~seen for an NS particle is shifted in mass and normalization.  For an S particle, $m_{\rm min} = 50\,{\rm GeV}$, and for $\mchi \geq 50\,{\rm GeV}$, $\sigv_{\rm min} = \sigvdm \approx 2\times 10^{-26}\,{\rm cm^{3}\,s^{-1}}$.  Note that when $\mchi \geq 100\,{\rm GeV}$, $\sigv_{\rm min,NS}/\sigv_{\rm min,S} = 2$.  In the mass range from 50 to 100 GeV, $\sigv_{\rm min,NS}$ increases approximately as $\mchi^{-1}$, while $\sigv_{\rm min,S}$ is approximately constant.  As a result, when $\mchi = 50$ GeV, $\sigv_{\rm min,NS}/\sigv_{\rm min,S} \approx 4$.  Thereafter, for $\mchi < 50\,{\rm GeV}$, the ratio $\sigv_{\rm min,NS}/\sigv_{\rm min,S} = 4$, remains constant for all lower masses.  As confirmed by Fig.\,\ref{fig:sigvlogall14b}, when allowing for $\omchi < \omdm$, there is no bound on the relic particle mass, although there is a lower bound to the annihilation cross section.

\begin{figure}
\begin{center}
\includegraphics[width=0.45\columnwidth]{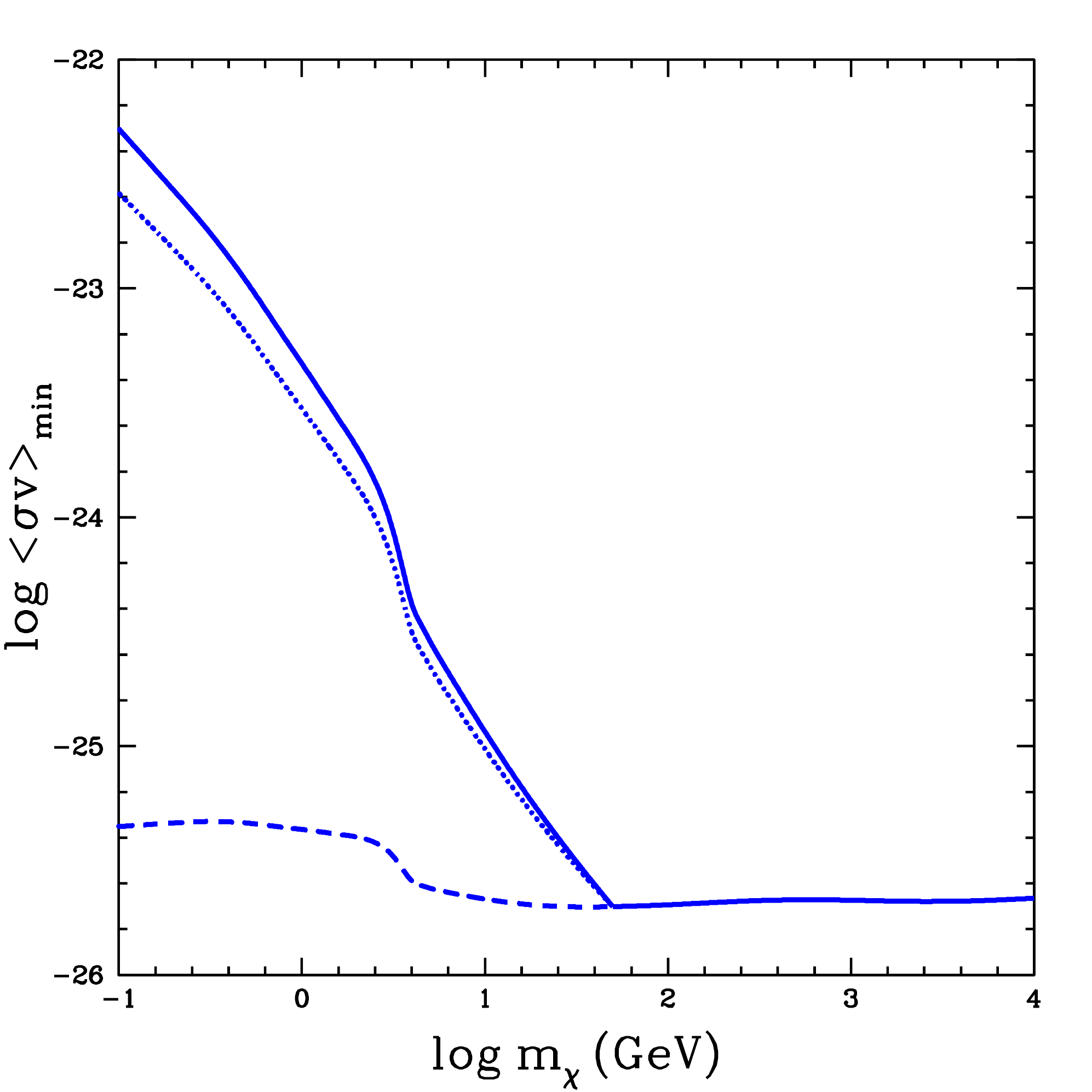}
\includegraphics[width=0.45\columnwidth]{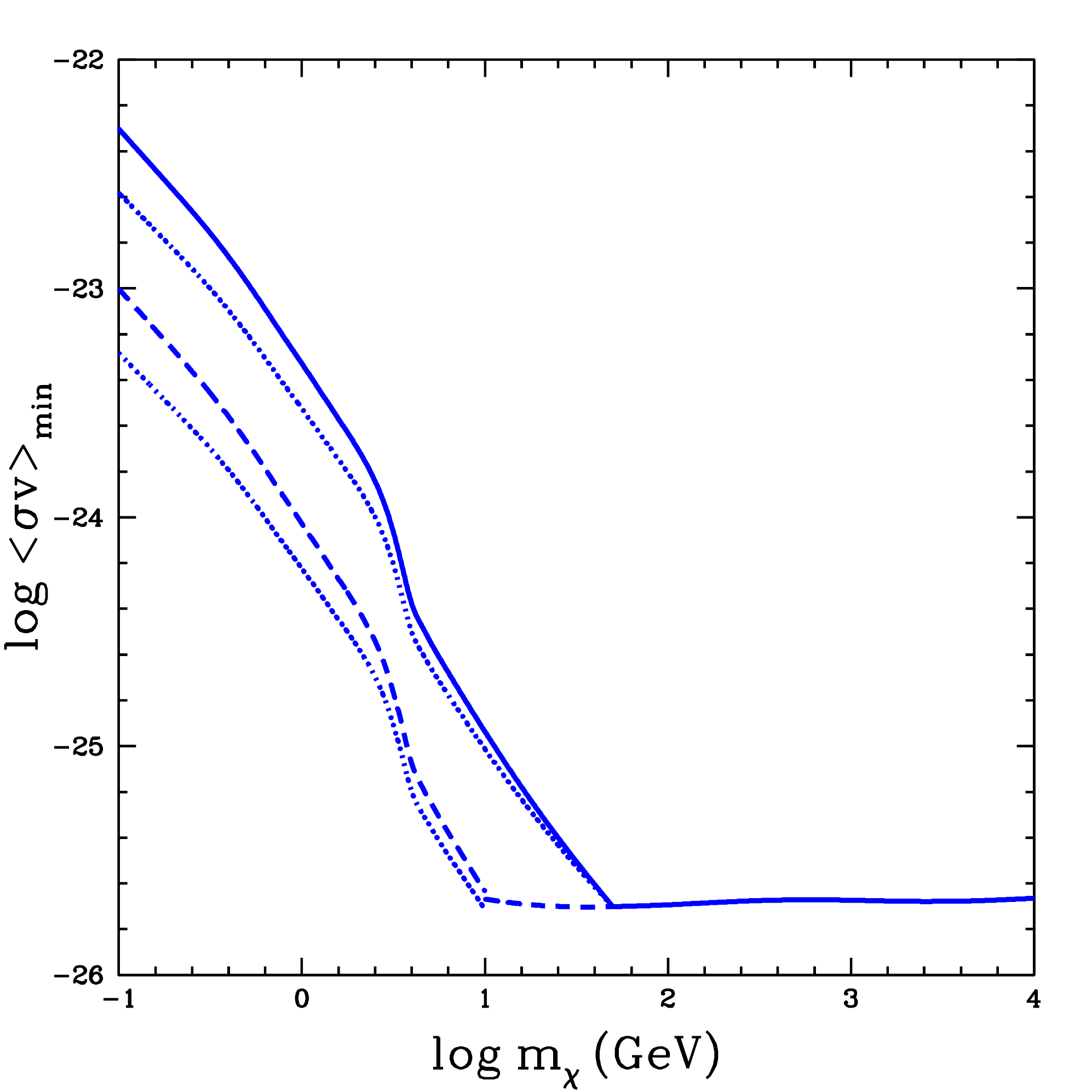}
\end{center}
\caption{(Color online) The two panels show the results for the minimum annihilation rate factor for a self-conjugate (S) thermal relic, $\sigv_{\rm min}$, satisfyin g the CMB constraint and $\omchi \leq \omdm$, as in Fig.\,\ref{fig:sigvlogall14b}. In both panels the dotted curves account for the logarithmic corrections to $\sigv_{\rm min}$ discussed in the text.  The regions above the solid (dotted) curves are allowed, while the regions below them are excluded.  In the left hand panel the solid (dotted) curve is for all masses and for $f = 1$.  The dashed curve shows the extension of $\sigv_{\rm DM}$ to lower masses, $m_{\chi} < m_{\rm min}$, illustrating that $\sigv_{\rm min} > \sigvdm$ in this mass range.  The right hand panel compares $\sigv_{\rm min}$ for $f = 1$ (solid) and $f = 0.2$ (dashed). }
\label{fig:sigvmlogtest}
\end{figure}

The results outlined here and shown in Fig.\,\ref{fig:sigvlogall14b} are expanded upon in the two panels of Figure \ref{fig:sigvmlogtest} where, for an S particle, the corresponding $\sigv_{\rm min}$ curves are shown as functions of the WIMP mass, including the logarithmic corrections from Eq.\,\ref{eq:chivsdm} that were ignored in Fig.\,\ref{fig:sigvlogall14b}.  In the left hand panel, $\sigv_{\rm min}$ and \sigvdm~are compared for $f = 1$.  In the right hand panel, the results for $\sigv_{\rm min}$ are shown for $f = 0.2$ as well as for $f = 1$.

\begin{figure}[!t]
\begin{center}
\includegraphics[width=0.45\columnwidth]{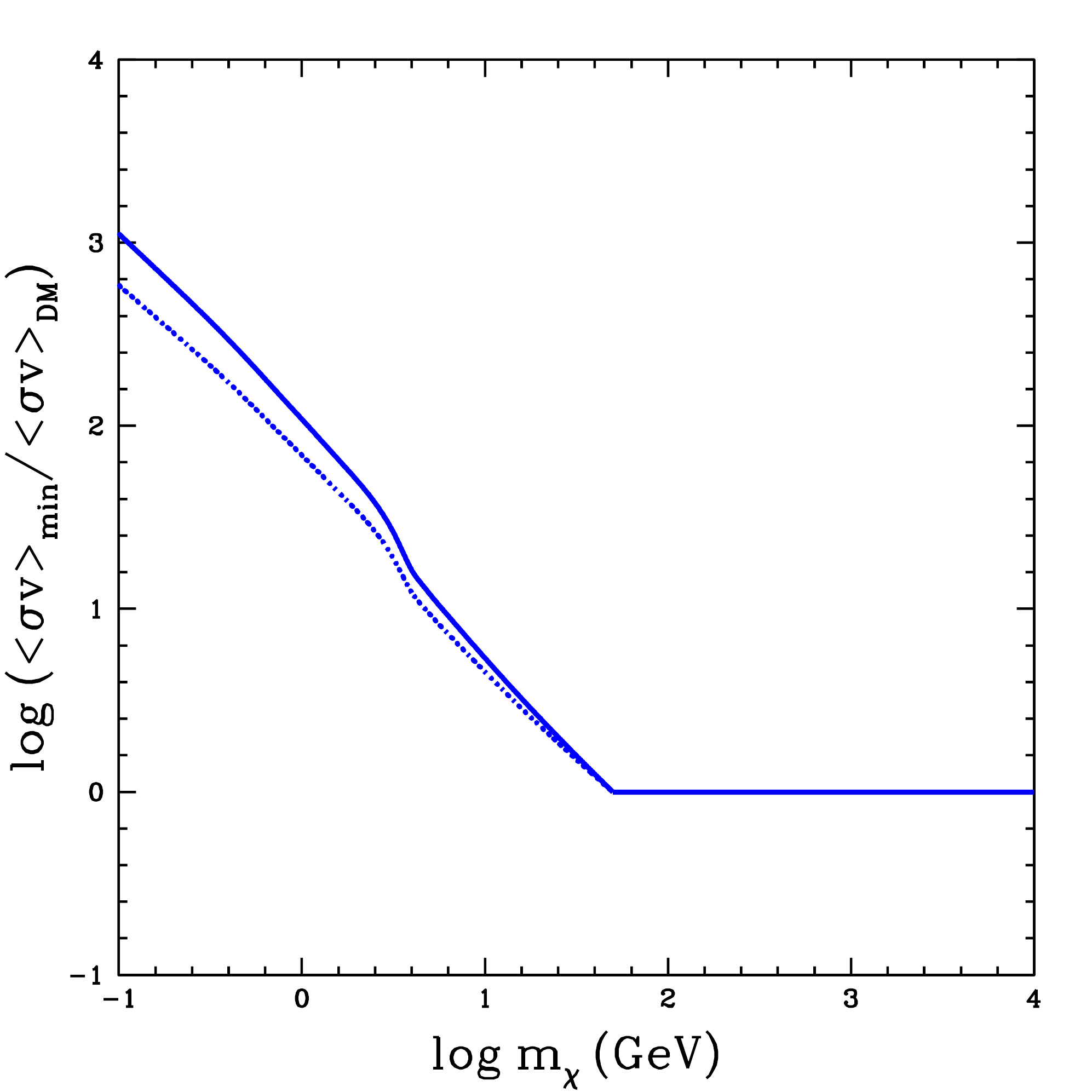}
\hskip .3in
\includegraphics[width=0.45\columnwidth]{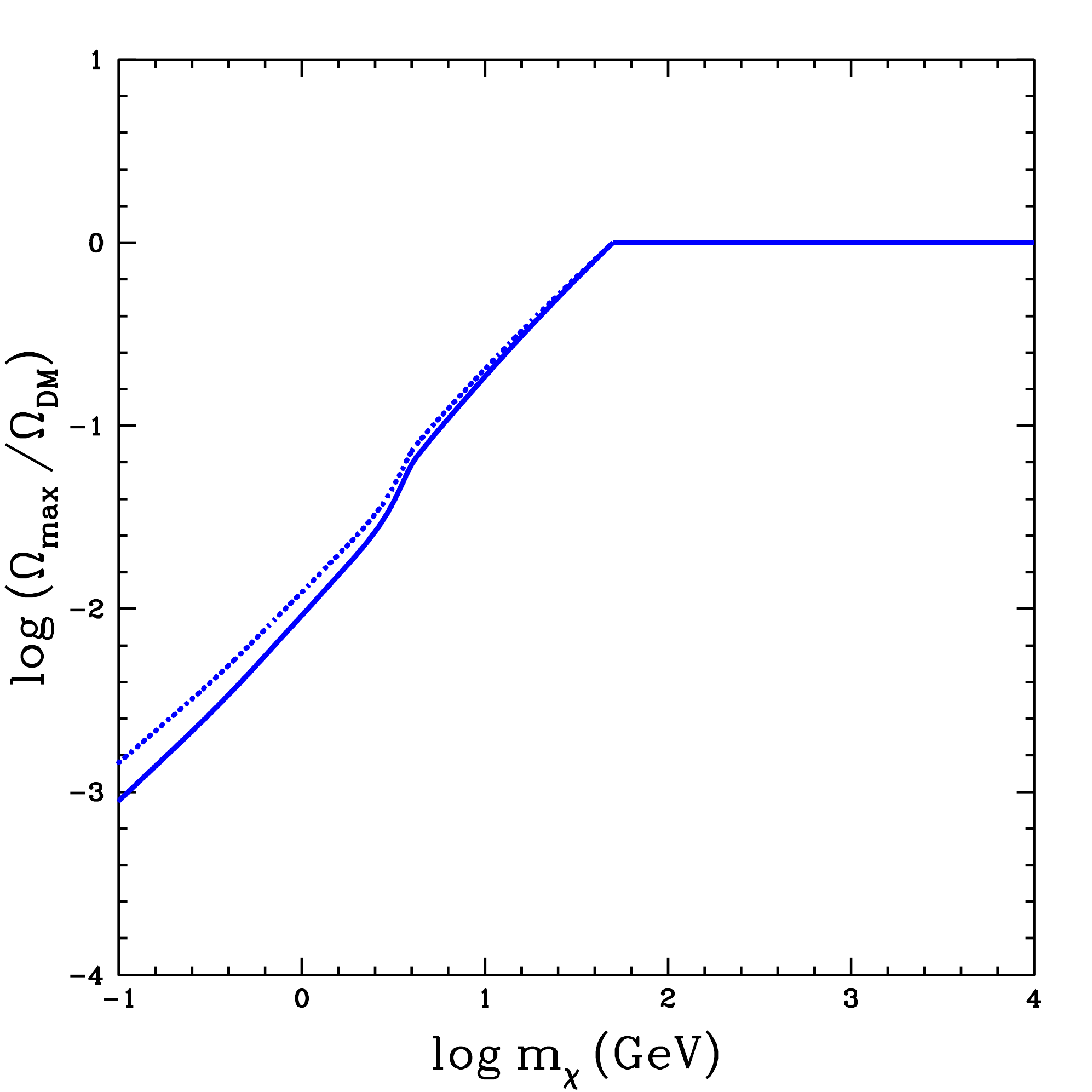}
\\\vskip 0.2in
\caption{(Color online) The left hand panel shows (for $f = 1$) the ratio of the minimum annihilation rate factor to the DM annihilation rate factor, $\sigv_{\rm min}/\sigvdm$, as a function of the WIMP mass.  The regions above the curves, corresponding to $\sigvchi \geq \sigv_{\rm min} \geq
\sigvdm$, are allowed, while the regions below the curves are excluded.  The right hand panel shows the variation of the ratio of mass densities, $\Omega_{\rm max}/\Omega_{\rm DM}$, with WIMP mass (also for $f = 1$).  For the ratio of mass densities, the allowed regions are restricted to lie below the curves ($\omchi \leq \Omega_{\rm max} \leq \omdm$).  In both panels the dotted curves account for the logarithmic corrections to $\sigv_{\rm min}$ discussed in the text.}
\label{fig:sigvratio}
\end{center}
\end{figure}

Including the previously neglected logarithmic corrections results in
\beq
\sigv_{\rm min} \approx {(\sigvdm)^{2} \over \sigvcmb}\,\bigg[1 + {1 \over 8}\,{\rm log}\,\bigg({\sigvdm \over \sigvcmb}\bigg)\bigg]^{-2} \lsim {(\sigvdm)^{2} \over \sigvcmb}\,,
\eeq
or
\beq
{\sigv_{\rm min} \over \sigvdm}  \approx {\sigvdm \over \sigvcmb}\,\bigg[1 + {1 \over 8}\,{\rm log}\,\bigg({\sigvdm \over \sigvcmb}\bigg)\bigg]^{-2} \lsim {\sigvdm \over \sigvcmb}\,.
\eeq
For $f = 1$ the ratio of cross sections, $\sigv_{\rm min}/\sigvdm$, is shown in the left hand panel of Figure \ref{fig:sigvratio}.  The regions {\bf above} the curves ($\sigvchi \geq \sigv_{\rm min}$) are allowed, consistent with $\omchi \leq \omdm$ and with the CMB constraint on late time annihilations.  

When $\sigvchi \geq \sigv_{\rm min}$, there is an upper limit to the relic mass density, $\omchi \leq \Omega_{\rm max} \leq \omdm$, where
\beq
{\Omega_{\rm max} \over \omdm} \approx {\sigvcmb \over \sigvdm}\,\bigg[1 + {1 \over 8}\,{\rm log}\,\bigg({\sigvdm \over \sigvcmb}\bigg)\bigg] \lsim {\sigvcmb \over \sigvdm}\,.
\eeq
This ratio is shown for an S particle, for $f = 1$, as a function of the relic particle mass in the right hand panel of Figure \ref{fig:sigvratio}.  Here, the regions {\bf below} the curves ($\omchi  \leq \Omega_{\rm max}$) are allowed.

\section{Summary And Discussion}
\label{sec:concl}

Although a thermal relic particle freezes out during the early evolution of the Universe, it doesn't cease annihilating; ``annihilations are forever" \cite{fkt}.  At freeze out, when $T = T_{f}$, the annihilation rate (per particle) is equal to the expansion rate, $\Gamma_{{\rm ann},f} = H_{f}$.  Throughout the subsequent evolution of the Universe, for $T < T_{f}$, $\Gamma_{\rm ann} < H$, and annihilations do continue, albeit at an ever decreasing rate ($\Gamma_{\rm ann} \ll H$).  However, since the energy released in each annihilation, 2\mchi, can be very large compared to the thermal energy (the temperature, $T < T_{f} \ll \mchi$) of the universal plasma, the energy released by the electromagnetic component of these rare, late time WIMP annihilations may leave an imprint on the CMB frequency or power spectrum \cite{msw,padman,galli09,galli11,hutsi,slatyer}.  Indeed, the current, best constraints are inferred from observations of the CMB power spectrum \cite{planck15}.  The CMB constraint, $\sigvcmb \leq 4\times 10^{-28}\mchi/f\,{\rm cm^{3}\,s^{-1}}$, when compared to the WIMP annihilation cross section required if the WIMP is a DM candidate (\omchi~= \omdm), results in setting a lower bound to the mass of the DM WIMP (see Fig.\,\ref{fig:sigvallchi14bb}).  For a stable, symmetric ($n_{\chi} = n_{\bar{\chi}}$), WIMP whose annihilation is s-wave dominated, $m_{\rm min} = 50\,(10)\,{\rm GeV}$ for $f = 1\,(0.2)$ if the particle is identical to its antiparticle (S: $\chi = \bar{\chi}$), and $m_{\rm min} = 100\,(20)\,{\rm GeV}$ for $f = 1\,(0.2)$ if the particle differs from its antiparticle (NS: $\chi \neq \bar{\chi}$).  Any further reduction in \sigvcmb~from future CMB experiments would increase $m_{\rm min}$.  For example, if the cosmic variance limit \cite{galli09}, a factor of four below the current Planck result \cite{planck15} were reached, the lower bound on the DM mass would increase by a factor of four compared to the current constraints.

These CMB constraints on WIMP DM candidates change dramatically if the WIMP is {\bf not} a dark matter candidate, but only contributes a fraction of the DM mass density ($\omchi < \omdm$, $\sigvchi > \sigvdm$).  As seen in \S\,\ref{sec:ndm} and illustrated in Figs.\,\ref{fig:sigvlogall14b}\,-\ref{fig:sigvratio}, in this case there is no bound to the WIMP  mass but, there is a {\bf lower} bound to the annihilation cross section, $\sigv_{\rm min}$, set by a combination of \sigvdm~and \sigvcmb.  Although all masses are allowed, for $\mchi \geq m_{\rm min}$, $\sigv_{\rm min} = \sigvdm$, while for $\mchi < m_{\rm min}$, $\sigv_{\rm min} = (\sigvdm)^{2}/\sigvcmb > \sigvdm$.  The lower bound to the annihilation cross section corresponds to an upper bound to the relic mass density, $\omchi \leq \Omega_{\rm max}$, where, up to logarithmic corrections, $\omchi/\Omega_{\rm max} \approx \sigv_{\rm min}/\sigvchi$ (see the right hand panel of Fig.\,\ref{fig:sigvratio}).

To illustrate the potential importance of the discussion here, consider the following application.  There is interest in constraining the electric charge of a fractionally charged particle ($q = Qe$).  A search for millicharged particles at SLAC \cite{slac}, led to upper bounds on $Q$ for particles with masses in the range from 100 keV to 100 MeV.  For $m_{Q} \lsim 100\,{\rm MeV} < m_{\mu}$, annihilations can only lead to \epm pairs or to photons.  For $m_{Q} \gsim m_{e}$ and $Q \ll 1$, annihilation to \epm pairs dominates, with an annihilation rate factor, $\sigv_{Q} \approx 2\times 10^{-21}Q^{2}/m_{Q}^{2}\,({\rm cm^{3}\,s^{-1}})$, where $m_{Q}$ is measured in GeV.  At a mass of 100 MeV, the SLAC experiment set an upper bound to the charge of $Q \leq 5.8\times 10 ^{-4}$.  Saturating this maximum charge, leads to an upper bound to the annihilation cross section, $\sigv_{Q} \lsim 6.7\times 10^{-26}\,({\rm cm^{3}\,s^{-1}})$.  Comparing this upper bound to the CMB lower bound to $\sigv_{Q} \geq \sigv_{\rm min} = 1.8\times 10 ^{-22}$ (at \mchi~= 0.1 GeV) shown in Fig.\,\ref{fig:sigvallchi14bb}, reveals a strong inconsistency.  The CMB constraint on the annihilation cross section, in combination with the SLAC upper bound to the electric charge, eliminates the possibility of a 100 MeV millicharged particle.  Indeed, extending the results for $\sigv_{\rm min}$ presented here to lower masses and repeating this comparison, rules out any fractionally charged particle in the mass range from 100 keV to 100 MeV, complementing the constraints on higher mass fractionally charged particles ($\sim {\rm few\,GeV}$) presented in \cite{ls}.  

Before concluding, it is worth commenting on a recent preprint \cite{kamion14}, in which Blum, Cui, and Kamionkowski also relax the assumption that the WIMP is a dark matter candidate, allowing for $\omchi \leq \omdm$.  The authors consider several observational consequences of late time annihilations, including the effect on the CMB.  However, since they restrict their attention to larger WIMP masses than those investigated here, $\mchi \geq 100\,{\rm GeV} \geq m_{\rm min}$, they do not consider the effects of the CMB constraint from late time annihilations on the lower bound (or not) of the WIMP mass and the lower bound to the annihilation cross section discussed here.  As noted in \cite{kamion14} and here, the results for $\omchi < \omdm$ have consequences for predictions of the expected gamma ray flux from late time annihilations in the present Universe in, \eg, the Galaxy, dwarf galaxies, or the intergalactic medium in clusters of galaxies.  In calculating the expected gamma ray fluxes, it is almost always assumed that the WIMP accounts for all of the dark matter in the astronomical target of interest (\ie, $n_{\chi} = n_{\rm DM}$).  If, however, $n_{\chi} < n_{\rm DM}$, the predicted flux must be rescaled (reduced) by a factor of $(\omchi/\omdm)^{2} < 1$.  In addition, the annihilation cross section must also be rescaled (increased) by a factor of $\sigvchi/\sigvdm \approx \omdm/\omchi > 1$.  The overall effect is to reduce the expected gamma ray flux by a factor of $(\omchi/\omdm)$ or, accounting for the logarithmic correction, by a factor of $(\omchi/\omdm)[1 - {\rm log}\,(\omchi/\omdm)/8]$.  The upper bound to this flux ratio, $\Omega_{\rm max}/\omdm$, is shown in the right hand panel of Fig.\,\ref{fig:sigvratio}.

The analysis presented here has avoided the case of an asymmetry between particles and antiparticles ($n_{\chi} \neq n_{\bar{\chi}}$) since for this case the relic abundance of the dominant particle (antiparticle) depends on the adopted asymmetry and not directly on the annihilation cross section.  In the presence of an asymmetry, since the relic abundance of the subdominant antiparticle (particle) is suppressed by continued annihilations after the dominant particle has frozen out ($n_{\bar{\chi}} \ll n_{\chi}$), the energy injection from the late time annihilations ($p_{\rm ann} \propto n_{\chi}\,n_{\bar{\chi}}$) is also suppressed.  As a result, it is not unlikely that an asymmetric WIMP could account for all of the dark matter, while being immune to the CMB constraint on late time annihilations (\eg, the lower bound to the WIMP mass could be much smaller than the value(s) of $m_{\rm min}$ derived here).  For recent discussions of asymmetric dark matter, see, \eg, \cite{adm1,adm2,adm3}.

In summary, it has been shown here that if a WIMP $\chi$, a thermal relic, accounts for all of the dark matter ($\omchi = \omdm$), the CMB constraint on electromagnetic energy injection from late time WIMP annihilations sets a lower bound to its mass, $\mchi \geq m_{\rm min}$.  For the current CMB data \cite{planck15}, this lower bound is $m_{\rm min} = 50\,(100)\,{\rm GeV}$, for S (NS) particles and an electromagnetic energy efficiency factor $f = 1$ (for $m_{\rm min} \gsim 10\,{\rm GeV}$, $m_{\rm min}$ scales linearly with $f \leq 1$).  However, if the WIMP only accounts for a part of the dark matter, there is no limit to its mass, but there is a lower bound to its annihilation cross section, $\sigvann \geq \sigv_{\rm min}$, that increases inversely with the WIMP mass since, $\sigv_{\rm min} = (\sigvdm)^{2}/\sigvcmb \propto f/\mchi$.

\acknowledgments

The author is grateful to the Ohio State University Center for Cosmology and Astro-Particle Physics for support of this research.  The genesis of this work was in conversations with N.\,Padmanabhan and I thank him for helpful explanations.  I am grateful to J.\,Beacom for many valuable comments and suggestions.  I also thank P.\,Gondolo for comparing the output from the DarkSUSY code with the results quoted in \S\,{\ref{sec:sigv} and shown in Figs.\,\ref{fig:sigvdm14}\,\& \ref{fig:sigvallchi14bb}.

\end{document}